# Decision Evidence Maturity Model for Agentic AI: A Property-Level Method Specification

Oleg Solozobov [1]

## Abstract

Agentic AI systems produce decision evidence at scale through execution telemetry, but property-level reconstruction often fails when an external party asks a specific governance question about a specific decision: the assembled evidence is insufficient to answer it. We name this pattern the container fallacy: the automatic equation of evidence-container presence with audit sufficiency. This paper specifies the Decision Evidence Maturity Model (DEMM), a property-level reconstructability method for agentic decisions. DEMM classifies evidence sufficiency into four executable categories plus a protocol-level *conflicting* category and aggregates per-property verdicts into a five-level capability rubric anchored to the established maturity-model lineage. The open-source Decision Trace Reconstructor ships ten executable adapter-fallback classes spanning vendor SDKs, protocol traces, public-postmortem prose, and generic JSONL records. A reproducible feasibility exercise runs the protocol on 140 synthetic scenarios plus three public incidents; the resulting completeness range (53.6 % to 100 %) is implementation behaviour, not external validation.

# 1. Introduction

Agentic AI systems — LLM-driven planners that chain tool calls, multi-agent orchestrators that delegate subtasks, and human-in-the-loop pipelines — are moving into production deployments where decisions carry real consequences. The 2025-2026 AI governance and agentic AI auditability response has been extensive: AI inventories, model cards, drift monitors, runtime kill-switches, audit ledgers, risk-management frameworks, and a fast-growing preprint literature proposing evidence containers (provenance graphs, signed delegation tokens, lifecycle ledgers, tool-firewall logs, statistical text watermarks, forensic artefact recovery) to render agentic execution legible after the fact. Mechanisms multiplied. Yet a recurring pattern surfaced across the most-discussed failure cases: when an external party asks a specific question about a specific decision, the assembled evidence is frequently *insufficient* to answer it, even when every individual mechanism functioned as designed.

[1] *Corresponding author.* Affiliation: Independent Researcher (Global). E-mail address: dev404ai@gmail.com. ORCID: https://orcid.org/0009-0009-0105-7459.

We name this pattern *mechanisms without sufficiency* and isolate the inferential failure — *automatic equation of container presence with audit sufficiency* — as the **container fallacy**: an inferential step that treats evidence-container presence as if it implied audit sufficiency, when sufficiency requires per-question per-property checking that container presence does not perform. The diagnostic is positioned above broad auditability dimensions (Nian et al., 2026a) and the *attestability* dimension of the four-axis ODTA (*observability/decidability/timeliness/attestability*) runtime-placement test (Koch & Wellbrock, 2026), not as a replacement for either.

The diagnostic resolves into a sufficiency predicate evaluated *per governance question, per decision-evidence property, over whatever evidence regime persisted*: a question an external party is entitled to ask, a property class drawn from the Decision Event Schema (Solozobov, 2026b), and the actually-persisted evidence regime jointly determine sufficiency. The auditability target is ODTA Attestability rather than generic observability (Koch & Wellbrock, 2026). The formal ternary relation $S \subseteq \mathcal{Q} \times \mathcal{P} \times 2^{\mathcal{E}}$ that operationalises this predicate, the eight-to-seven property collapse, and the $singleton/cross-regime$ composition rules are stated in §3.1. Two operationalisations follow: a property-level sufficiency check (this paper) and an *Overclaim Rate* measurement under controlled degradation (forthcoming benchmark paper).

This paper specifies the **Decision Evidence Maturity Model (DEMM)**, a property-level method that operationalises governance-evidence sufficiency for an automated decision as reconstructability over a finite property schema, discrete sufficiency categories, and a five-level maturity rubric. The schema is the *Decision Event Schema* (MIT) (Solozobov, 2026b); the reference implementation is the *Decision Trace Reconstructor* ($Apache-2.0$) (Solozobov, 2026a), which exercises it on synthetic and named-incident traces to show protocol feasibility where container-presence baselines do not. The evaluation is deliberately framed as feasibility, not independent empirical validation.

## 1.1. The Gap This Paper Fills

Three production agentic-AI incidents from 2025-2026 make the gap operationally concrete (full reconstructions in §4.1 and §5.4): the Replit DROP DATABASE event of July 2025 (McGregor & Database, 2025), the Cursor destructive-shell-command incident of August 2025 (Cursor forum (user-reported), 2025), and the Claude Code DataTalks.Club Terraform-destroy event of February 2026 (McGregor & Database, 2026). In each, the destructive *action* is documented but the auditor's question – *under what governance regime was this destructive mutation authorised?* – is not answered by any single record on its own.

The synthesis paper on the three structural breaks of agentic governance (Solozobov, 2026e) sits upstream. The accountability canon establishes reconstruction-capable instruments as a precondition for accountability, each entry identifying the gap between *evidence existing* and *evidence answering the question.* Moral crumple zones (Elish, 2019) and accountable algorithms (Kroll et al., 2017) frame the diffusion-of-blame and procedural-evidence questions. Administrative-state accountability (Busuioc, 2021) and the monitorability problem (Yampolskiy, 2024) extend the same gap to bureaucratic and AI-monitor settings. Auditable Agents (Nian et al., 2026a) defines the broad auditability framework; the present paper contributes a narrower property-level sufficiency layer above it.

## 1.2. Research Questions

The present paper addresses three research questions. **RQ1 (primary):** Can governance-evidence sufficiency for an agentic decision be operationalised as per-property reconstructability over a finite decision-evidence property schema, and classified into discrete sufficiency categories? **RQ2 (secondary):** Can an open-source reference implementation exercise the per-property check reproducibly across heterogeneous evidence regimes on synthetic scenarios and named-incident reconstructions, while exposing where container-presence baselines overclaim? **RQ3 (tertiary):** Can the per-property classification be aggregated into a five-level maturity progression – *ad-hoc* to *process-attested* to *property-instrumented* to *sufficiency-tested* to *continuously-attested* – compatible with the established maturity-model lineage (full anchors in §3.7)?

## 1.3. Contributions

Three contributions follow from the research questions. **(i) Property-level operationalisation of Beyond Task Success ODTA Attestability** (Koch & Wellbrock, 2026): per-property sufficiency over the decision-event schema (Solozobov, 2026b), framing inherited from the present authors' precursor on evidence sufficiency under delayed ground truth (Solozobov, 2026d). **(ii) Five-level DEMM rubric** anchored to the established maturity-model lineage of the Capability Maturity Model and its successors (CMM and CMMI), the Test Maturity Model integration (TMMi), and the Data Management Capability Assessment Model (DCAM); the substrate (per-property evidence sufficiency) is the contribution, the level pattern is borrowed. **(iii) Decision Trace Reconstructor** reference implementation + reproducible feasibility exercise on 140 synthetic scenarios plus three named-incident reconstructions across a 3 × 2 architecture × stack-coverage matrix; the cross-regime adapter tier is specified at the API level but not yet fully executable, stated openly rather than asserted as a future-work footnote.

The broad auditability-framework coordinate is owned by Auditable Agents (Nian et al., 2026a); the present contribution is a property-level sufficiency layer above auditability dimensions and above evidence containers. The §5 reconstructability tensor is a feasibility output from the deposited Decision Trace Reconstructor (Solozobov, 2026a), not a calibrated population estimate or an independent validation dataset; full quantitative validation under controlled degradation is the subject of the forthcoming benchmark paper.

## 1.4. Structure of the Paper

§2 surveys evidence containers and the *evidence regime* construct; §3 specifies DEMM (pipeline, categories, tensor, five-level rubric); §4 fixes the 3 × 2 matrix, three named-incident reconstructions, three metrics, and one comparator; §5 reports per-cell reconstructability, boundary accuracy, the **instrumentation premium**, and **container-baseline overclaim**; §6 reads results against prior theory, threats to validity, diagnostic implications for audit-surface boundaries, and a **five-axis differentiation triangle**; §7 closes with future work, limitations, broader-impact.

# 2. Background and Related Work

This section positions §3 against agentic-governance literature, surveys seven evidence-container classes, defines *evidence regime*, and names the upper-layer frames DEMM operationalises or sits above.

## 2.1. Agent Observability Is Not Governance Evidence

Contemporary agent-observability tools — OpenTelemetry GenAI, LangSmith, AWS Bedrock AgentCore, OpenAI Agents SDK, Anthropic Managed Agents — provide execution telemetry in detail but are not governance-oriented: traces support execution reconstruction but do not answer authority-grade sufficiency questions about a specific decision. Tracing SDKs emit spans per tool call; observability platforms index for search and alerting; orchestration telemetry records which agent called which tool. None asks, per decision-event property, whether that property is filled, unfillable, or opaque by design. Trace-Based Assurance (TBA) names the gap: "standard logs capture events but rarely provide semantic accountability: which role made which claim, based on which evidence, under which constraints, and with what verification outcome" (Paduraru et al., 2026). Execution observability is necessary substrate, not sufficient for accountability.

## 2.2. Seven Heterogeneous Container Classes — Survey

The 2025-2026 agentic-governance literature catalogues seven heterogeneous evidence-container classes, each supporting one audit question while leaving an adjacent decision-evidence property unanswered. The first three classes are runtime and provenance records: agent execution records (Vispute & Kadam, 2026), workflow provenance graphs (Souza et al., 2025), and lifecycle audit trails (Ojewale et al., 2026). The next two cover authorisation and presentation: signed delegation tokens, instantiated as DCC (Patil, 2026) and HDP (Dalugoda, 2026), and final-text statistical signals (Nian et al., 2026b). The last two cover mediation and forensics: tool-firewall logs (Yuan et al., 2026a) and recovered local artefacts (Gruber & Hilgert, 2026). AER supports reasoning-provenance questions but not organisational principal authority; *DCC*/*HDP* support signed delegation but not downstream state mutation; AEGIS-NTC supports *allow*/*block* boundary questions but not target-system mutation after a permitted call. These examples supply the substrate §5.6's container-baseline overclaim measures against; the same seven classes appear in §3.2 as upstream evidence regimes.

The *signed delegation token* class merges DCC and HDP as twin instantiations of the same authorisation envelope. Both cede state-mutation questions (Patil, 2026; Dalugoda, 2026). Every surveyed container leaves at least one property class unanswered, and the unanswered question is in each case *adjacent* to one the container answers well. Adjacency is what makes the fallacy persuasive: if a firewall log captures a permitted call and the call was destructive, the natural reading is that the log "covers" the destruction, when the firewall instruments the boundary and is silent about the target (Yuan et al., 2026a).

Six anchors position DEMM.

**Beyond Task Success** (Koch & Wellbrock, 2026) introduces the four-axis ODTA runtime-placement test (*observability/decidability/timeliness/attestability*) plus the MAEB (minimum action-evidence bundle for state-changing actions) within a four-layer *Evaluation/Governance/Orchestration/Assurance* framework. DEMM operationalises ODTA Attestability at the property level over MAEB-compatible records.

**AER** (Vispute & Kadam, 2026) — contemporaneous reasoning provenance verified via transport-layer interceptors. DEMM classifies reasoning trace as *opaque* (§3.4); AER is an upstream regime the §3.2 adapter tier consumes.

**TBA** (Paduraru et al., 2026) — four-layer prospective assurance with a runtime mediator. The F1, F2, and F3 failure classes co-describe the three structural breaks (Solozobov, 2026e); §6.7 reads the convergence. MAT records deserialise directly into DEMM pipeline fragments.

**Runtime Governance** (Kaptein et al., 2026) — formalises compliance policies over agent identity, path, action, and state. It populates Orchestration; DEMM populates Assurance. *Feasibility precedes enforcement* (§6.4) is the boundary.

**Parallax** (Fokou, 2026) — *cognitive/executive* separation with boundary validator. Parallax classifies admissibility; DEMM classifies sufficiency. IFL propagation is T8; F8 closes.

**AgentCity** (Ruan & Zhang, 2026a) — multi-principal agent economies via SoP. DEMM is single-principal; cross-principal extension is F5.

## 2.3. Evidence Regimes as a Cross-Cutting Construct

Surveyed frameworks produce records that differ along four orthogonal properties — *producer role*, *temporal mode*, *cooperation assumption*, and *schema commitment*. AER's interceptor model (Vispute & Kadam, 2026), TBA's runtime mediator (Paduraru et al., 2026), and OpenClaw's forensic recovery (Gruber & Hilgert, 2026) anchor the cooperative-runtime, prospective-runtime, and adversarial-recovery extremes respectively. Public post-mortems (McGregor & Database, 2026; McGregor & Database, 2025; Cursor forum (user-reported), 2025) anchor the non-cooperative-narrative extreme. Property-level reconstructability must operate over these coordinates rather than framework names; this paper introduces the *evidence regime* construct as the four-property coordinate. A single framework may emit records in multiple regimes.

The §3.2 adapter tier covers eighteen evidence-regime classes in this coordinate; per-class executable conformance is split between currently-executable and protocol-level scope, with the partition specified in §3.2.

The four-property coordinate enables **cross-regime assessment** (same protocol across distinct regime coordinates) and **regime substitutability**: when two regimes occupy compatible coordinates and supply the same property fragments, the per-property verdict is invariant. IEEC chains fill the same authorisation-envelope properties as DCC tokens (He & Yu, 2026; Patil, 2026). Subsequent uses of "regime" / "cross-regime" refer to this four-property coordinate and the adapter contract in the deposited artefact (Solozobov, 2026a).

### 2.4. Auditable Agents and the Three-Layer Reading

The novelty defence is **three-layered**: Auditable Agents (Nian et al., 2026a) defines auditability dimensions, mechanism classes, layered evidence, and Auditability Card; PROV-AGENT, lifecycle audit-ledger work, OSCAL-style compliance, AER, TBA, IEEC, and memory-runtime graphs define the substrates; the present method measures whether records produced under any of those regimes suffice for property-level reconstruction. The paper cedes broad-framework priority to Auditable Agents. On the Auditable Agents $detect/enforce/recover$ mechanism partition (Nian et al., 2026a), DEMM positions an explicit **assess** coordinate alongside the verify and present coordinates that adjacent $runtime/presentation$ work occupies; the verify, present, and assess coordinates extend the Auditable Agents three-class partition rather than being inherited from it. Full landscape positioning is developed in §6.8.

## 3. DEMM Method Specification

This section specifies the **Decision Evidence Maturity Model (DEMM)** method, the load-bearing technical contribution of the paper. DEMM is a property-level reconstructability assessment for single-principal agentic deployments; the reference implementation is the *Decision Trace Reconstructor* (DTR; specification in §1).

The method's job is to project records drawn from heterogeneous evidence regimes onto a common decision-event property schema – the *Decision Event Schema* (MIT) (Solozobov, 2026b) – and report, per property and per chain, whether the available evidence is sufficient for a specific post-hoc governance question. Within Beyond Task Success's four-layer framework (Koch & Wellbrock, 2026), DEMM operationalises the ODTA Attestability class at the property level: it answers neither whether to enforce nor how to enforce, but whether the evidence that was in fact persisted, recovered, or reconstructed from postmortem narratives suffices to attest to a governance question after execution.

## 3.1. Design Goals, Non-Goals, and the Sufficiency Relation

DEMM operates strictly post-hoc on stored or recovered evidence rather than instrumenting the decision-making process in real time, and bounds scope to single-principal deployments under a non-cooperative-by-default assumption: it works with whatever evidence the upstream regimes produced, partially produced, or failed to produce, and asks only whether that evidence suffices for the post-hoc governance question at the property level. DEMM is invoked on a closed artefact set (agent execution logs, tool-call records, model outputs, policy snapshots, schema-formatted fragments, adapter-normalised regime records) and emits per-property sufficiency verdicts after the fact. Non-goals: DEMM is not a debugger, does not produce evidence (upstream regimes do), does not enforce at runtime ($TBA/Parallax/Runtime$ Governance do), does not establish legal compliance, and does not generalise to multi-principal economies (F5 of §7). When evidence is insufficient, DEMM reports a typed feasibility verdict (§3.5) rather than synthesising a plausible-looking chain.

**Formal sufficiency relation.** Auditability of an agentic decision is a ternary relation $S \subseteq \mathcal{Q} \times \mathcal{P} \times 2^{\mathcal{E}}$ over (i) a set $\mathcal{Q}$ of governance questions an external party is entitled to ask, (ii) a set $\mathcal{P}$ of decision-evidence properties, and (iii) a set $\mathcal{E}$ of evidence regimes surveyed in §2.3. The eight property classes — actor identity, principal authority, action boundary, policy basis, decision basis, data and resource touch, lifecycle context, verification strength — are derived from the six DES schema field groups (Solozobov, 2026b) and refined into the agentic-decision granularity required by per-property reconstructability; this eight-class refinement is the present paper's contribution layered above the DES schema, not a verbatim enumeration in the cited source. The v0.1.0 reference implementation collapses *actor identity* and *principal authority* into a single row, yielding the seven-property metric used in §5. Atomic regimes $E \in \mathcal{E}$ identify with their singleton assemblies $\{E\} \in 2^{\mathcal{E}}$, so $S(Q, P, E)$ is read as $S(Q, P, \{E\})$; cross-regime compositions $\mathcal{E}' = \{E_1, \dots, E_k\} \in 2^{\mathcal{E}}$ admit the same predicate. Sufficiency is the audit-relevant predicate, evaluated per question and per property over whatever evidence persisted. The auditability target is ODTA Attestability rather than generic observability (Koch & Wellbrock, 2026).

## 3.2. Evidence-Regime Adapters and Reconstruction Pipeline

The assessment layer is a two-tier construct: an upstream **evidence-regime adapter tier** that normalises records from cooperative and non-cooperative regimes into a shared fragment family, and a downstream **six-stage reconstruction pipeline** — fragment collection, temporal ordering, chain assembly, decision boundary detection, decision-event schema mapping, and feasibility report generation — that operates on the normalised fragments and emits the per-property sufficiency verdicts of §3.5; v0.1.0 realises the adapter tier as ten executable adapter-fallback classes spanning vendor SDKs, protocol traces, public-postmortem prose, and generic JSONL records, while eight further regime classes remain protocol-level conformance work. Each stage carries an input contract, output contract, feasibility gate, and failure verdict; Table 1 summarises the contract surface.

**Table 1. Stage-contract surface.** Per stage, the input object the stage consumes, the output object it emits, the feasibility gate that conditions stage entry, and the failure verdict the stage emits when the gate is unmet. Contracts are realised by the deposited release (Solozobov, 2026a).

| Stage | Input | Output | Feasibility gate | Failure verdict |
|---|---|---|---|---|
| 0 — Adapter | Regime-native records (per §2.3 regime class) | Typed fragment stream over §3.2 fragment family | Adapter declared for the source regime class | *regime_unsupported* (*structurally_ unfillable* for all properties downstream) |
| 1 — Fragment collection | Typed fragment stream | Typed-pointer manifest with lazy content resolution | Manifest non-empty | *no_fragments_ recovered* (*structurally_ unfillable* per property) |

| Stage | Input | Output | Feasibility gate | Failure verdict |
|---|---|---|---|---|
| 2 — Temporal ordering | Typed-pointer manifest | Causally ordered fragment stream | Timestamps or explicit causal edges resolvable | *unorderable_fragments* (*partially_fillable* with order-confidence < threshold) |
| 3 — Chain assembly | Causally ordered fragments | Candidate decision chains, attributed | Attribution recoverable on each fragment | *unattributed_evidence_rejected* (per-fragment verdict) |
| 4 — Boundary detection | Candidate chains | Partitioned decision units with confidence scores | At least one of four §3.3 heuristics fires | *boundary_unrecoverable* (*partially_fillable* with low confidence) |
| 5 — Schema mapping | Decision units | Per-property fillability flags over $\mathcal{P}$ | Property class has a defined fragment-to-property mapping | per-property *structurally_unfillable* (annotated with architectural reason) |
| 6 — Feasibility report | Per-property flags | Per-property + per-chain category (§3.5) + recommendation tensor (§3.6) | All upstream stages emitted a verdict (success or typed failure) | gate-log itself becomes evidence about the audited system |

**Adapter tier (Stage 0).** Per regime, the adapter declares fragment-family mapping (decision-event log, tool-call payload, policy snapshot, authorisation chain, reasoning placeholder, post-condition state), identifier mapping, and which properties can in principle be filled. Ten primary adapter-fallback classes with worked examples ship in the deposited release: IEEC (He & Yu, 2026); $DCC/HDP$ (Patil, 2026; Dalugoda, 2026); AER (Vispute & Kadam, 2026); TBA MAT (Paduraru et al., 2026); Springdrift runtime DAG (Brady, 2026); LanG $UICR/MCP$ audit (Abdennebi et al., 2026); AEGIS-NTC tool-firewall (Yuan et al., 2026a); OpenClaw artefact-only (Gruber & Hilgert, 2026); public-postmortem prose driving §5.4. Eight further regimes named in §2.3 are protocol-level only; executable conformance is deferred to the benchmark paper. Adapter-normalised fragments enter the pipeline as a single typed stream.

**Stage notes beyond Table 1.** Stage 2 inherits logical-clock discipline (Lamport, 1978) and the Dapper distributed-tracing tradition (Sigelman et al., 2010), realised over OpenTelemetry GenAI parent-trace IDs where available. Stage 4 heuristics are detailed in §3.3 and Stage 5 mapping fillability in §3.4-§3.5. The gate log emitted by every stage is itself evidence about the audited system: a stream of typed verdicts that downstream auditors and the §3.6 instrumentation recommendation tensor consume jointly with the per-property tensor.

## 3.3. Decision Boundary Detection within Stage 4

The reconstructor detects decision boundaries in continuous agent flows using four configurable heuristics — state-change magnitude (flags state-deviation thresholds), tool-call boundaries (every invocation externalises internal reasoning), human-intervention points (explicit handoffs treated as confident), policy-constraint activation (constraint-firing signals different regime) — and reports a confidence score per placement; the heuristic status is principled because no ground truth exists for boundary location in continuous flows.

Boundary detection is a $Stage-4$ component, not a headline novelty: it makes boundary identification a first-class output rather than an assumption from the tracing tool. Confidence reporting allows downstream controls to condition on confidence; §5.2 shows which cells the four-heuristic composition handles well in the deposited artefact (Solozobov, 2026a).

## 3.4. Authorization Envelope Reconstruction within Stage 5

For agent steps where internal reasoning is opaque (LLM generation events are canonical), the reconstructor does not attempt internal-logic reconstruction but reconstructs the authorisation envelope: inputs available (prompt, tool outputs, retrieval), constraints active (policies, tool-permission scopes, model limits), and outputs possible (tool calls allowed, format constraints, terminal action set) given system configuration.

The construction respects the ML-opacity boundary (Solozobov, 2026c): post-hoc reconstruction of internal reasoning conflates governance with explainability. Reasoning trace is therefore classified *opaque* uniformly across the deposited suite (LLM-only, HITL-supervised, and non-agentic rule-based steps all reduce to the authorisation envelope at this property level). §3.5 treats *opaque* at weight 1.0 because the substituted envelope is the actionable governance fragment; what varies across §5 cells is envelope richness, not the reasoning-trace classification. The §3.2 adapter tier determines that richness — $DCC/IEEC$ regimes supply full signed chains; weaker regimes supply partial envelopes with explicit gap descriptions. Proof-carrying authorisation (Cohen, 2026) is the formal-verification adjacent supplying the strongest fragment-class input under cooperative regimes.

### 3.5. Feasibility Categories

The assessment layer classifies each decision-event property and each reconstructed decision into one of five sufficiency categories grounded in the present authors' precursor on evidence sufficiency under delayed ground truth (Solozobov, 2026d) — the precursor four-dimension evidence-sufficiency framing (completeness, freshness, reliability, representativeness) is extended here with five fillability categories specific to agentic-decision reconstruction: **fully fillable**, **partially fillable**, **structurally unfillable**, **opaque** (ML-opacity boundary; envelope substituted per §3.4), **conflicting** (cross-regime fragments inconsistent). The first four are exercised by §5; *conflicting* is protocol-level with worked examples — executable conflict-scoring and precedence rules are near-term future work (§7, F4). Each verdict carries actionable payload: *fully_fillable* the reconstructed value; *partially_fillable* recoverable evidence + gap description; *structurally_unfillable* no value but annotated with architectural reason (cross-stack boundary, state lost, evidence never persisted, non-cooperative stripping); *opaque* the authorisation envelope; *conflicting* the disagreeing fragment values + source regimes + placeholder for the conflict-resolution policy.

Reconstruction completeness is a weighted average completeness$(r) = \sum_{p \in \mathcal{P}} w_r(p)/|\mathcal{P}|$, $|\mathcal{P}| = 7$ per the implementation schema (Solozobov, 2026a). Weights: *fully_fillable = 1.0*; *opaque = 1.0* (envelope-substituted per §3.4); *partially_fillable = confidence in [0, 1]* (default *0.5* in v0.1.0 — not independently calibrated, so downstream cross-regime comparisons are protocol-relative; calibration on independent traces is the forthcoming benchmark target); *structurally_unfillable = 0.0*. The category is a decision-event field, making per-architecture failure modes legible at the cell level: each cell is characterised by which category dominates in which property class.

### 3.6. Output Tensor and Instrumentation Recommendations

The assessment-layer output is a structured tensor over architecture × stack-coverage × evidence regime × property × sufficiency category, accompanied by an instrumentation recommendation naming the upstream regime that would close each missing or partial property. §5 reports a sparse slice: a fully populated $3 \times 2 \times 7 \times 4$ synthetic matrix plus a public-postmortem regime slice instantiated for three single-agent cross-stack incidents (§5.4). The v0.1.0 reconstructor emits *gap-closing* recommendations for *partially_fillable* and *structurally_unfillable* properties, *optional enrichment* recommendations for *opaque* properties (which already carry the §3.4 envelope at weight 1.0). The recommendation rule is regime-specific rather than generic: policy and authorisation gaps route to explicit execution-contract or delegation records, *action/state* gaps route to runtime or firewall records, cross-stack gaps route to trace-context propagation, and non-cooperative recovery routes to forensic artefacts. Recommendations for the protocol-level *conflicting* category are deferred to the executable extension (§7, F4).

The mapping is implemented in the reference artefact (Solozobov, 2026a); the resulting tensor + recommendation operationalises §6.4's *safe-claim envelope*: a deployment cannot defensibly claim attestability for properties classified *structurally_unfillable* or *conflicting* until the recommended upstream regime is in place. This is *feasibility precedes enforcement* (§6.4) made operational, with ODTA Attestability the bounding target (Koch & Wellbrock, 2026).

## 3.7. The Five-Level DEMM Maturity Rubric

The Decision Evidence Maturity Model organises governance-evidence sufficiency along a five-level capability progression – *ad-hoc* to *process-attested* to *property-instrumented* to *sufficiency-tested* to *continuously-attested* – that mirrors the established maturity-model lineage ($CMM/CMMI$, TMMi, DCAM); the unit of capability at each level is the *evidence the deployment can produce, on demand, for a specific governance question about a specific decision* – not the maturity of the surrounding governance organisation, the standardisation of the evidence interchange format, or the posture of the telemetry platform. The five levels are descriptive of the evidence regime the deployment carries into post-hoc audit, and a deployment is assigned a level on a per-property-class basis; the framing rule applies throughout this subsection.

The five-level pattern inherits the $CMMI/TMMi/DCAM$ maturity-lineage form (Paulk et al., 1993; TMMi Foundation, 2018; EDM Council, 2020), with the substrate changed to per-property evidence sufficiency. 1. **Ad-hoc.** Reactive manual assembly on challenge; turnaround days-weeks. 2. **Process-attested.** Governance processes (model cards, control documentation, committees); manual reconstruction; turnaround hours-days. 3. **Property-instrumented.** Runtime captures schema-named properties by design; reconstruction automated; turnaround minutes-hours. 4. **Sufficiency-tested.** Property-instrumented regime exercised against a question battery; per-question $pass/fail$ under $S(Q, P, E)$. 5. **Continuously-attested.** Sufficiency monitored as SLO; degradation triggers alarms; posture maintained across $retraining/drift/handoff/migration$.

**Lineage anchors.** The rubric inherits the *progression + level + capability-substrate* triple from data-analytics maturity literature (Langer, 2025); adjacent AI-governance instantiations supply context without replacing the per-property substrate (Gasser & Almeida, 2017; Shneiderman, 2020; Janssen et al., 2020). Levels 1-5 mirror the $CMMI/TMMi/DCAM$ “ad-hoc to continuously instrumented” pattern (Paulk et al., 1993; TMMi Foundation, 2018; EDM Council, 2020). The SR 26-2 vocabulary mapping (§6.4) is a later regulatory reading of the substrate, not a source of the level definitions (Federal Reserve System; OCC; FDIC, 2026). Substrate is the contribution; level pattern is borrowed.

**Per-property aggregation.** A deployment is assigned a level per property-class (e.g., Level 4 for *actor identity* + Level 2 for *decision basis* simultaneously). The aggregate level is the **minimum across the property classes the forum’s governance questions actually test** — avoiding the standard maturity-model failure mode where a high aggregate conceals a critical gap. Threshold values inside a level are proposed design parameters; the forthcoming benchmark paper provides empirical calibration. DEMM is distinct from explainability-maturity rubrics MM4XAI-AE (Muñoz-Ordóñez et al., 2025) and provenance-based interpretability verification (Vonderhaar et al., 2026).

# 4. Evaluation Design

The evaluation is a controlled feasibility exercise: the reconstructor is deployed against decision chains whose ground truth is either synthetic-generator-derived or public-reporting-documented; per-property reconstructability, boundary-detection accuracy, and dominant-break attribution are measured across a $3 \times 2$ matrix. The design is pre-specified and committed to the deposited release (§4.4), whose citable artefact fixes the release tag, seed list, per-cell manifests, result tables, and bootstrap scripts. The numbers test reproducible execution of the specification, not population rates. Full *Overclaim Rate* measurement under controlled degradation is the forthcoming benchmark paper.

## 4.1. Experimental Setup

Primary evaluation data are 140 synthetic agentic scenarios deposited on Zenodo (Solozobov, 2026a); named-incident reconstructions are secondary illustrative evidence under the *public record establishes / paper infers* convention. The ordering reflects pre-enrolment scope: production traces from industrial partners are excluded (no IRB pre-enrolment, proprietary traces cannot be shared). Synthetic scenarios are reproducible protocol fixtures, not sampled field traces; named incidents anchor them to ecologically valid failures without converting them into population evidence.

**The *public record establishes / paper infers* convention.** §5.4 reconstructions are based on public reporting. Each named incident reports in two clauses: *Public record establishes* — what the public reporting documents (cited to $URL/artefact + retrieval$ date); *paper infers, for evaluation* — what DEMM's per-property protocol reads off the public record. Adapts digital-forensics chain-of-custody (Kent et al., 2006; $ISO/IEC$ JTC 1/$SC$ 27, 2012) + blameless-postmortem SRE culture (Beyer et al., 2016).

Three named incidents are admitted under this convention with stratified evidentiary strength.

The **Claude Code DataTalks.Club incident** (February 2026) — AIID-curated entry. *Public record establishes:* the AIID curated record (McGregor & Database, 2026) documents a Claude Code session in which a Terraform destroy command wiped the DataTalks.Club course-platform infrastructure ($database$+2.5 years of student submissions and snapshots); the operator restored from AWS Business Support backups. *Paper infers:* the session exercises the property classes the §5 single-agent cross-stack cell fails to fill on synthetic chains, and the auditor's "under what governance regime was this authorised?" surfaces the same per-property gap. Architectural context catalogued in a recent design-space analysis (Liu et al., 2026); full reconstruction in §5.4.

The **Replit DROP DATABASE incident** (July 2025) — vendor-grade with official-mitigation caveat. *Public record establishes:* a Replit coding-assistant destructively dropped a production database in an authorised session; Replit's official response (McGregor & Database, 2025) catalogued platform mitigations (rollback windows, scope-locked permissions, deploy guards, updated default-allow policies). *Paper infers:* the mitigation catalogue is itself diagnostic — each closes a property-level gap the original session left open. We read the set not as Replit endorsing this paper's framing but as engineering judgement that the original evidence regime was insufficient at the property level. Full reconstruction in §5.4.

The **Cursor user-forum report** — caveated user-level pattern evidence. *Public record establishes:* a forum thread (Cursor forum (user-reported), 2025) documents a Cursor coding agent issuing a destructive shell command (*rm -rf ~*) in an active development session. *Paper infers:* the same per-property gap recurs across IDE-coupled coding agents, but the report is anonymous and user-level rather than vendor-grade — community-level pattern evidence, not a closed forensic case.

For each cell of the 3 × 2 matrix (§4.2), the reconstructor is run against twenty synthetic scenarios selected to cover representative tool topologies and decision depths; this $n = 20$ is a fixture-coverage design choice, not a power calculation or precision claim. The synthetic data generator is deterministic given a seed, so the twenty scenarios per cell are reproducible bit-for-bit from the deposited release (Solozobov, 2026a). The scenario design exercises the Decision Event Schema property set (Solozobov, 2026b) against the three structural breaks specified in the upstream taxonomy (Solozobov, 2026e). The 120 agentic scenarios across the 3 × 2 matrix plus 20 non-agentic baseline scenarios constitute the 140 scenario feasibility exercise reported in §5.

## 4.2. Architecture × Stack-Coverage Matrix

The evaluation matrix is a 3 × 2 cell grid: three agentic architectures (single-agent tool-use, multi-agent orchestration, and human-in-the-loop agentic) crossed with two stack-coverage tiers (within-stack governed runtimes and cross-stack boundaries); a non-agentic baseline anchors the comparison. The architecture axis operationalises the distinct decision topologies of contemporary agent deployments. Single-agent tool-use captures the modal LLM-agent pattern: one agent, a fixed toolset, and a sequence of tool calls leading to an outcome. Multi-agent orchestration captures *planner/executor* and manager-worker patterns in which multiple agents exchange messages and delegate actions. Human-in-the-loop agentic captures the supervised pattern in which an agent proposes actions that a human approves, rejects, or edits before execution.

The stack-coverage axis distinguishes within-stack runs (entirely inside a governed runtime emitting structured telemetry — AgentCore, LangGraph+LangSmith, Anthropic Managed Agents, OpenAI Agents SDK) from cross-stack runs (crossing runtime boundaries — MCP servers, browser tools, mutable retrieval sources). The 3 × 2 matrix yields six cells; the non-agentic baseline (classical discrete-decision system with explicit policy execution) anchors the comparison and the upper bound on completeness (Solozobov, 2026a).

Table 2 gives the resulting cell grid and the expected dominant structural break per cell.

**Table 2.** Evaluation matrix: architectures × stack-coverage tiers with representative systems per cell and expected dominant structural break.

| Architecture | Within-stack (governed runtime) | Cross-stack (boundary spans) |
|---|---|---|
| Single-agent tool-use | AgentCore / LangGraph+LangSmith; expected dominant break: **evidence fragmentation** (tool-output capture) | Agent + external MCP servers / browser tools; expected: **evidence fragmentation** (amplified) |
| Multi-agent orchestration | LangGraph multi-agent / OpenAI Agents SDK; expected: **decision diffusion** | MCP-mediated multi-agent; expected: **decision diffusion** (severe) |
| Human-in-the-loop agentic | Anthropic Managed Agents with approval gates; expected: **responsibility ambiguity** | HITL with external tool execution; expected: **responsibility ambiguity + fragmentation** |
| Non-agentic baseline (anchor) | Discrete rule+model system with explicit decision point | — |

## 4.3. Metrics

The evaluation pre-specifies three primary metrics plus one RQ2 comparator. (1) **Reconstruction completeness** = weighted average of per-property contributions per chain (formula: completeness($r$) $= \sum_{p \in \mathcal{P}} w_r(p)/|\mathcal{P}|$, with $|\mathcal{P}| = 7$; weights per §3.5), reported $mean + 95\%$ bootstrap CI over the fixed $n = 20$ fixture set per cell. (2) **Boundary-detection accuracy** = F1 vs ground-truth boundaries on synthetic chains (suppressed on named incidents). (3) **Seven-mode failure dominance** = modal operational failure mode + % share among unrecoverable properties per cell. The secondary comparator is the **container-presence overclaim indicator**: each §2.2 container class is scored *sufficient* when present and well-formed, then compared with Table 7 property verdicts; Rows 1-6 report strict pp overclaim; Row 7 qualitative. Metric implementation, bootstrap seeds, and raw per-cell result tables are deposited with the reference artefact (Solozobov, 2026a). Validity is tiered: boundary F1 uses synthetic ground truth; completeness, per-property verdicts, seven-mode dominance, and overclaim indicators are mapper outputs, not independent labels. Completeness is the primary cross-cell outcome; boundary F1 separates boundary error from evidence gap; seven-mode dominance maps to severity. The seven-mode rubric extends the three structural breaks of the present authors' synthesis paper (Solozobov, 2026e): **Mode 1** *prompt/context* loss; **Mode 2** delegation-chain fragmentation; **Mode 3** channel gap; **Mode 4** schema mismatch; **Mode 5** policy-snapshot absence; **Mode 6** implicit-policy coupling; **Mode 7** authorship ambiguity. Grouping: decision $diffusion = modes$ {1,2,5,6}, evidence fragmentation = {3,4}, responsibility ambiguity = {7}. Thus *structurally_unfillable* means unfillable under the published mapper and artefacts, not impossible from extra-record knowledge.

### 4.4. Reproducibility

All synthetic scenarios, the Decision Trace Reconstructor release tag, pre-specification metadata, per-cell JSON manifests, raw feasibility reports, per-cell result tables, bootstrap seeds, and evaluation scripts are deposited as a citable artefact (Solozobov, 2026a). IRB approval is not required (synthetic data and publicly reported incidents only). The artefact is the pre-specification anchor for the evaluation contract: release tag **v0.1.0** of the deposited artefact (Solozobov, 2026a) pins the commit hash, deposit date, seed list, per-cell JSON manifests, raw feasibility reports, and per-property confidence scores at the version-of-record level; an independent replicator with Python 3.11 can regenerate every §5 table and figure from that pinned release.

Synthetic scenarios are bitwise-reproducible from seeds; named-incident reconstructions are not (public reporting may update), so the Zenodo release pins reporting URLs retrieved on a specified date with retrieved artefacts attached (Solozobov, 2026a).

## 5. Results

The reconstructor was run on 140 scenarios (120 agentic across the $3 \times 2$ matrix, 20 per cell; 20 non-agentic baseline). Named-incident reconstruction covers the Replit DROP DATABASE event of July 2025 (McGregor & Database, 2025), the Cursor destructive-command event of August 2025 (Cursor forum (user-reported), 2025), and the Claude Code DataTalks.Club Terraform-destroy event of February 2026 (McGregor & Database, 2026). All numbers are reproducible bit-for-bit from the pinned seed list of the deposited release (Solozobov, 2026a) and should be read as fixture-suite protocol outputs under the specified mapper, not as independent labels or population estimates.

### 5.1. Reconstruction Completeness by Cell

Within the synthetic fixture suite, reconstruction completeness varies systematically across the $3 \times 2$ matrix, with the non-agentic baseline at ceiling (100.0%) and multi-agent cross-stack lowest (53.6%); the architecture axis exhibits the consistent ordering HITL $>$ single-agent $>$ multi-agent in both within-stack and cross-stack tiers. Per-cell completeness is reported in Table 3.

**Table 3.** Reconstruction completeness by cell (mean % with 95% bootstrap CI, $n = 20$ fixture scenarios per cell). These are fixture-suite protocol outputs under the published mapper. Formula and per-category weights per §3.5 / §4.3. Reasoning trace is uniformly *opaque* across all 140 scenarios and contributes 1.0 to every cell; cell-to-cell variation reflects the other six properties. Empty cell at right: baseline runs entirely within stack by construction.

| Architecture | Within-stack | Cross-stack |
|---|---|---|
| Single-agent tool-use | 85.0 [83.6, 85.7] | 71.1 [70.4, 71.4] |
| Multi-agent orchestration | 67.5 [62.5, 72.1] | 53.6 [50.7, 57.1] |
| Human-in-the-loop agentic | 94.3 [94.3, 94.3] | 79.3 [78.2, 80.0] |
| Non-agentic baseline | 100.0 [100.0, 100.0] | — |

The agentic cells average 24.9 pp below the non-agentic baseline (100.0%). Under the pre-specified matrix, this decomposes descriptively into a 17.7 pp architecture cost plus a 7.2 pp cross-stack contribution; within-stack agentic mean is 82.3%, cross-stack mean 68.0%. The largest single-cell gap is multi-agent cross-stack at 53.6%, 46.4 pp below baseline. Substantive interpretation against RQ1 and the three-break taxonomy is deferred to §6.1.

## 5.2. Boundary-Detection Accuracy

Boundary-detection F1 (Table 4) is suppressed for the baseline and named incidents (no ground truth). The striking result is multi-agent within-stack F1 = 0.13 under the current four-signal heuristic and the synthetic mapper's authored downgrade rules. Multi-agent cross-stack records 0.46. Single-agent (0.84/0.76) and HITL (0.73/0.74) F1 scores track completeness. The raw boundary labels, detector outputs, and bootstrap setup are deposited with the Decision Trace Reconstructor artefact (Solozobov, 2026a); the upstream taxonomy reference is (Solozobov, 2026e). Substantive interpretation of the multi-agent F1 collapse and of the 0.46 cross-stack reading is deferred to §6.1; threshold sensitivity is recorded in T12 (§6.5).

**Table 4.** Boundary-detection F1 on synthetic chains (mean over 20 scenarios per cell, tolerance +/-1 fragment). Fixture means only: this detector diagnostic checks authored labels, not the Table 3 completeness estimator. $N/A$ for non-agentic baseline and for named-incident scenarios.

| Architecture | Within-stack | Cross-stack |
|---|---|---|
| Single-agent tool-use | 0.84 | 0.76 |

| Architecture | Within-stack | Cross-stack |
|---|---|---|
| Multi-agent orchestration | 0.13 | 0.46 |
| Human-in-the-loop agentic | 0.73 | 0.74 |

## 5.3. Seven-Mode Failure Dominance and the Severity Ranking

The hypothesised severity ranking — decision diffusion first, evidence fragmentation second, responsibility ambiguity third — is hypothesis-consistent in the synthetic data: multi-agent cells (architecture-axis decision-diffusion drivers; cross-stack additionally fragmentation-amplified per Table 5) lose the most completeness against baseline (mean 40 pp), single-agent cells (fragmentation) lose intermediate (mean 22 pp), HITL cells (ambiguity) lose the least (mean 13 pp). The ranking is not independently confirmed: the generator and mapper encode the hypothesised mechanisms, so the numbers measure expressive consistency rather than external validity (empirical confirmation on independently-sourced traces is F1 of §7). Modal seven-mode failures per cell appear in Table 5.

**Table 5.** Modal seven-mode failure per cell, share of unrecoverable properties, dominant break (§4.3). The Share denominator is per-cell *structurally_unfillable* property count over 20 scenarios. HITL within-stack reports *n/a* for modal mode and share because the responsibility-ambiguity break surfaces as *partially_fillable* of actor identity (Table 7: 52.1%), not as an unrecoverable property; the break is recorded in the Dominant-break column.

| Architecture | Tier | Modal mode | Share | Dominant break |
|---|---|---|---|---|
| Single-agent tool-use | Within-stack | Mode 4 (schema mismatch) | 1.00 | Evidence fragmentation |
| Single-agent tool-use | Cross-stack | Mode 3 (channel gap) | 1.00 | Evidence fragmentation |
| Multi-agent orchestration | Within-stack | Mode 6 (implicit policy) | 0.53 | Decision diffusion |
| Multi-agent orchestration | Cross-stack | Mode 3 (channel gap) | 1.00 | Evidence fragmentation |
| Human-in-the-loop agentic | Within-stack | n/a (no unrecoverable properties; see caption) | n/a | Responsibility ambiguity (surfaces as *partially_fillable*, not unrecoverable) |
| Human-in-the-loop agentic | Cross-stack | Mode 3 (channel gap) | 1.00 | Evidence fragmentation |

A second pattern visible in Table 5 is that in every cross-stack cell the modal mode is mode 3 (channel gap) regardless of architecture. The mode labels and per-cell failure counts are part of the deposited result tables (Solozobov, 2026a). Interpretation of the cross-stack-as-amplifier reading and its refinement of the three-break taxonomy (Solozobov, 2026e) is in §6.1.

## 5.4. Named-Incident Reconstruction Case Studies

Named-incident reconstruction of three public agentic incidents (Claude Code DataTalks.Club AIID record, Replit DROP DATABASE with mitigation catalogue, Cursor forum report) produces 57.1% weighted completeness — 14.0 pp lower than the 71.1% synthetic mean for their matching cell (single-agent cross-stack). We name this gap the **instrumentation premium**: the per-property weighted-completeness gap between a controlled synthetic-trace regime and a public-postmortem regime under the same DEMM protocol. Formally, with the Decision Event Schema property set (Solozobov, 2026b) and §3.5 / §4.3 weights, the premium = 71.1% − 57.1% = 14.0 pp at the single-agent cross-stack cell. The three incidents stratify by evidentiary strength under the §4.1 convention: Replit vendor-grade (CEO official response), Claude Code DataTalks.Club AIID-curated record, Cursor community-level pattern evidence. The premium is a small public-postmortem set within one matrix cell, not a regime-quality score in general.

Cases are selected under the §4.1 convention; the AI Incident Database (McGregor, 2020) is the population frame, not a sampled population.

All three cases reconstruct at 57.1% and share the Table 6 profile. **Claude Code DataTalks.Club** (McGregor & Database, 2026), **Replit DROP DATABASE** (McGregor & Database, 2025), and **Cursor user-forum report** (Cursor forum (user-reported), 2025) document $actor/action$ enough to fill actor identity and output action, but public records lack authorising $policy/tool-permission$ envelope; inputs and post-condition state remain partial, and reasoning remains opaque.

**Table 6.** Named-incident reconstruction feasibility summary. Seven-class implementation schema (Solozobov, 2026a); each row sums to 7 (2 $fully$ + 2 $partially$ + 2 structurally $unfillable$ + 1 opaque). *opaque* is the §3.4 reasoning-slot substitution.

| Incident | Architecture | Stack coverage | Fully fillable | Partially fillable | Structurally unfillable | Opaque |
|---|---|---|---|---|---|---|
| Claude Code DataTalks.Club (Feb 2026) | Single-agent | Cross-stack | 2 | 2 | 2 | 1 |
| Replit DROP DATABASE (Jul 2025) | Single-agent | Cross-stack | 2 | 2 | 2 | 1 |
| Cursor destructive shell command (August 2025) | Single-agent | Cross-stack | 2 | 2 | 2 | 1 |

Across all three incidents (McGregor & Database, 2026; McGregor & Database, 2025; Cursor forum (user-reported), 2025), Table 6's verdict pattern matches the deposited *named_incidents.json* (Solozobov, 2026a): **structurally_unfillable** = policy basis + action $boundary/configuration$ envelope; **opaque** = reasoning trace; **fully_fillable** = actor $identity/principal$ authority + output action; **partially_fillable** = inputs + post-condition state. Authorisation envelope is *fully_fillable* in synthetic scenarios (Table 7, snapshots present by construction) but *structurally_unfillable* in named incidents. With §3.5 weights, $(2 \cdot 1.0 + 2c + 2 \cdot 0 + 1 \cdot 1.0)/7$ and default $c = 0.5$ gives 57.1%, hence the 14.0 pp protocol-relative premium against the 71.1% synthetic mean. Substantive interpretation is in §6.1.

## 5.5. Per-Property Feasibility Matrices

Table 7 reports per-property feasibility across all 140 scenarios; deposited JSON supplies raw counts (Solozobov, 2026a), and the left column crosswalks the Decision Event Schema property set (Solozobov, 2026b) to v0.1.0 rows. Two endpoints are fixed by construction: action $boundary/configuration$ envelope is always *fully_fillable* where snapshots exist (bounded by T11), and reasoning trace is uniformly *opaque* (§3.4 substitution). Among contested properties, actor $identity/principal$ authority is the hardest.

**Table 7.** Per-property feasibility distribution across all 140 scenarios (% of scenarios; one-decimal rounding from *results/per_property.json*). The left column gives the $canonical - property/implementation - row$ crosswalk; v0.1.0 collapses actor identity and principal authority. Action $boundary/configuration$ envelope is 100% fully fillable where snapshots are present (T11); this implementation property is narrower than §3.4's *authorisation envelope*. Reasoning trace is uniformly *opaque* and contributes 1.0 to every cell's completeness numerator.

| DES conceptual property / v0.1.0 implementation row | Fully fillable | Partially fillable | Structurally unfillable | Opaque |
|---|---|---|---|---|
| data and resource touch / Inputs | 57.1 | 42.1 | 0.7 | 0.0 |
| policy basis / Policy basis | 72.9 | 0.0 | 27.1 | 0.0 |
| actor identity + principal authority / Actor identity and principal authority | 19.3 | 52.1 | 28.6 | 0.0 |
| action boundary / Action boundary and configuration envelope | 100.0 | 0.0 | 0.0 | 0.0 |
| decision basis / Reasoning trace with opaque substitution | 0.0 | 0.0 | 0.0 | 100.0 |
| verification strength / Output action | 76.4 | 0.0 | 23.6 | 0.0 |
| lifecycle context / Post-condition state | 54.3 | 41.4 | 4.3 | 0.0 |

The tensor concentrates structural unfillability in policy basis (27.1%), actor *identity/principal* authority (28.6%), and output action (23.6%). Inputs and post-condition state are dominated by *partially_fillable*. RQ1 is therefore answered at protocol-feasibility level: finite DES properties are classified into discrete sufficiency categories with per-property traceability. The tensor values come from the deposited per-property result artefacts (Solozobov, 2026a); interpretation against the three-break taxonomy (Solozobov, 2026e) is in §6.1.

## 5.6. Container-Presence Baseline Comparison: Systematic Overclaim

A container-presence baseline that scores by *presence* of named container classes (AER, workflow provenance, lifecycle audit trails, signed delegation tokens, final-text signals, tool-firewall logs, recovered artefacts) systematically overclaims sufficiency vs DEMM's per-property tensor; the overclaim concentrates at exactly the property classes each container's authors cede in §2.2. Each §2.2 class is evaluated as a baseline rule on the 140 scenario suite (*sufficient* when the container is present and well-formed, else *not_sufficient*); Table 7 is the per-property comparator. The contrast is sharpest for the property class each container cedes to adjacent regimes.

Table 8 reports the compact overclaim-indicator slice for all seven §2.2 container classes. The first six rows are quantitative indicators on the 140 scenario fixture suite; Row 7 is qualitative because OpenClaw's verification-strength comparator is scope-classified as *opaque-undefined* rather than

*partially_fillable*. Source anchors are AER (Vispute & Kadam, 2026), workflow provenance graphs (Souza et al., 2025), lifecycle audit trails (Ojewale et al., 2026), *DCC*/*HDP* signed delegation (Patil, 2026; Dalugoda, 2026), IET final-text signals (Nian et al., 2026b), AEGIS-NTC tool firewall (Yuan et al., 2026a), and OpenClaw artefact recovery (Gruber & Hilgert, 2026).

**Table 8.** Container-presence overclaim indicators. *Baseline* = *container* present and well-formed; DEMM *comparator* = *Table* 7 property verdict under the deposited mapper.

| # | Container class | Adjacent property tested | Baseline score | DEMM comparator | Indicator |
|---|---|---|---|---|---|
| 1 | AER | Actor identity / principal authority | 100% | 19.3% fully; 52.1% partial; 28.6% unfillable | +80.7 pp strict |
| 2 | Workflow provenance graphs | Policy basis | 100% | 72.9% fully; 27.1% unfillable | +27.1 pp |
| 3 | Lifecycle audit trails | Output action / per-run path | 100% | 76.4% fully; 23.6% unfillable | +23.6 pp |
| 4 | DCC / HDP signed delegation | Output action / downstream execution | 100% | 76.4% fully; 23.6% unfillable | +23.6 pp |
| 5 | IET final-text signals | Post-condition state | 100% | 54.3% fully; 41.4% partial; 4.3% unfillable | +45.7 pp strict |
| 6 | AEGIS-NTC tool firewall | Post-condition state / target mutation | 100% | 54.3% fully; 41.4% partial; 4.3% unfillable | +45.7 pp strict |
| 7 | OpenClaw artefact recovery | Verification-strength completeness | qualitative only | qualitative only | qualitative only |

Each quantitative row exhibits the adjacency §1's container fallacy identifies: the container instruments one side of the decision event, while the tested property lies in an adjacent regime.

These overclaim magnitudes share the T1 self-fulfilling-design dependency: the DEMM measurements against which the container-presence baseline is compared are produced by the deposited mapper (Solozobov, 2026a) authored to encode the same structural-break logic as the three-break taxonomy (Solozobov, 2026e). The *signed* and *adjacency-driven* pattern is less sensitive to that dependency because it follows from the literature's own "sufficient *for/not* sufficient for" framing; the specific pp magnitudes are feasibility measurements under this implementation, not population estimates. Independent-trace replication (§7 F1) and controlled-degradation *Overclaim Rate* measurement in the forthcoming benchmark paper are required before those magnitudes can be cited as empirical estimates.

**Row 7 qualitative.** OpenClaw (Gruber & Hilgert, 2026) under *artifact_only* yields no numeric overclaim entry: verification-strength is *opaque-undefined*, not *partially_fillable* (scope decision, §3.5). Container-presence still overclaims qualitatively by treating recovered artefacts as completeness wrt the audit question; qualitative overclaim = §2.2 Row 7 worked example.

The overclaim pattern in the six quantitative comparator rows is *signed* (every quantitative row overclaims; none under-claims) and *adjacency-driven* (each overclaimed property is adjacent to what the container actually instruments); Row 7 remains qualitative because its comparator is scope-classified as *opaque-undefined*. RQ2 is therefore answered at protocol-feasibility level: the same implementation runs across synthetic and named-incident regimes, and the container-presence comparator overclaims in every quantitative row. Values are DTR feasibility measurements (Solozobov, 2026a); full *Overclaim Rate* under controlled degradation is the benchmark paper. The vendor-claim-admissibility framework (Partasyuk, 2026) measures the same concern at the vendor-claim layer. Interpretation of this *signed/adjacency* pattern is in §6.1.

## 5.7. Null and Negative Results

Four *null/negative* results from the deposited release (Solozobov, 2026a). (i) **No *conflicting* outcomes** in 140 synthetic scenarios or three named incidents. (ii) **No detectable boundary signal** in multi-agent within-stack flows (F1 = 0.13, Table 4). (iii) **No across-incident divergence**: Table 6 returns the same profile for all three public-postmortem cases (McGregor & Database, 2026; McGregor & Database, 2025; Cursor forum (user-reported), 2025), an illustrative convergence, not population evidence. (iv) **No independent verdict adjudicator**: Table 7 categories are mapper outputs, not reconciler-validated labels. These baselines are recorded for downstream replication.

## 5.8. RQ3 Closure: Maturity Aggregation Output

RQ3 is answered by §5 as executable aggregation, not calibrated maturity validation: the DTR release emits the per-property tensor required by §3.7's minimum-over-tested-properties rule (Solozobov, 2026a). Table 7 is the bridge. *structurally_unfillable* properties cannot support property-instrumented or sufficiency-tested maturity for the tested question; *partially_fillable* properties remain below sufficiency-tested maturity until thresholds and verdict adjudication are calibrated; *opaque* decision-basis rows mature only at the substituted authorisation-envelope coordinate, not at explainability. Thus §5 demonstrates computability of the five-level rubric while withholding final deployment ratings until F7 calibration.

# 6. Discussion

§6.1 reads §5 against prior theory; §6.2 names principled scope; §6.3 reads diagnostic implications; §6.4 frames regulatory implications; §6.5 catalogues threats to validity; §6.6-§6.8 position DEMM against the surveyed landscape under the five-axis differentiation triangle.

## 6.1. Interpretation of §5: Feasibility Reading and Theory Refinement

The §5 tensor (Solozobov, 2026a) shows that the DEMM protocol can express the three-break taxonomy (Solozobov, 2026e) as per-property sufficiency differences. The architecture-axis ranking is hypothesis-consistent (multi-agent 40 $pp/single-agent\ 22/HITL$ 13), while the fixture design also exposes a cross-stack fragmentation amplifier — every cross-stack cell shows mode 3 (channel gap) dominance regardless of architecture (Table 5). The resulting refinement is *three architecture-native breaks with a cross-stack fragmentation amplifier whenever both are simultaneously active.* The §5.1 $baseline-\delta$ decomposition (24.9 *pp=17.7* pp $architecture+7.2$ pp cross-stack averaged across the matrix) is therefore a protocol output, not an external estimate. The HITL > single-agent > multi-agent ordering reflects explicit human approvals anchoring actor identity under HITL while multi-agent delegation chains diffuse those properties.

The multi-agent within-stack F1 = 0.13 (Table 4) is consistent with the decision-diffusion break (Solozobov, 2026e): the four-signal heuristic fires on almost every tool call when multiple agents delegate, so boundary localisation collapses in continuous multi-agent flows. The 0.46 cross-stack reading is better only because telemetry discontinuity removes spurious signals — the discontinuity is itself the fragmentation symptom (mode 3 dominance, Table 5). Single-agent (0.84/0.76) and HITL (0.73/0.74) F1 scores track completeness, so loss is driven by genuine evidence gaps. Stronger claims require independent traces, alternative detectors, or adversarial stress under the deposited release (Solozobov, 2026a) (F1 of §7); threshold sensitivity is T12.

The §5.5 overclaim pattern operationalises §1's container-fallacy diagnostic at the property level (container presence is not a sufficient predicate); the *signed* and *adjacency-driven* properties follow from the surveyed literature's own "sufficient *for/not* sufficient for" framing (§2.2), grounded in the broad-auditability frame (Nian et al., 2026a). The 14.0 pp instrumentation premium (§5.4) reads as a regularity within a shared reconstruction protocol rather than population-level evidence: the same seven-property recipe applied to three single-agent cross-stack postmortems yields the same residual gap. Confidence calibration is load-bearing for cross-comparison and is the target of the forthcoming benchmark paper (Solozobov, 2026a).

### 6.2. What DEMM Cannot Reconstruct — And Why That Is Principled

The ML-opacity boundary — substituting an authorisation envelope for opaque LLM reasoning rather than reconstructing internal logic — is a principled scope decision, not a limitation future versions should close. The grounding argument is in the present authors' prior work on governance evidence (Solozobov, 2026c): post-hoc reconstruction of internal neural reasoning conflates governance with explainability. The right question — *was the step inside its authorisation envelope?* — is what DEMM answers; *explain the LLM step* is deliberately abstained from. The same logic applies to §3.3 boundary detection: no ground truth for "where one decision ends" means boundary detection is classification with reported confidence, not recoverable truth.

### 6.3. Diagnostic Implications: Audit-Surface Boundaries

The §5 tensor exposes audit-surface boundaries — diagnostic readings about where governance evidence is and is not reconstructable — rather than a deployment recipe. Three boundary observations follow. (i) The matrix coordinate is itself a diagnostic: architecture × stack-coverage maps directly to §5.1 Table 3 reconstructability and §5.3 Table 5 dominant-break attribution. (ii) Structurally-unfillable concentration in policy basis (27.1%), actor *identity/principal* authority (28.6%), and output action (23.6%) marks the audit-surface boundary at the per-property level; the 0% structurally-unfillable reading on action *boundary/configuration* envelope is conditional on snapshot presence (T11), not a guarantee. (iii) The 14.0 pp instrumentation premium (§5.4) is a regime-comparison observation under the v0.1.0 partial-confidence default ($c = 0.5$), a warning indicator not a calibrated adjustment factor.

The multi-agent within-stack F1 = 0.13 (Table 4) marks the audit-surface edge for continuous multi-agent flows under the deposited four-signal heuristic (Solozobov, 2026a); T12 and F1 test whether threshold recalibration moves that edge.

The approximately 14.3 pp cross-stack penalty (§5.1) and mode 3 dominance (Table 5) diagnose trace-context discontinuity at runtime boundaries; whether instrumentation closes it is outside the post-hoc method.

The 28.6% structurally-unfillable actor-identity share (Table 7) marks evidence-granularity coupling: session-level attribution substitutes for step-level chains. The §3.6 $IEEC/DCC/HDP$ recommendation names the missing recovery regime.

## 6.4. Regulatory Framing — EU AI Act, SR 26-2, MAS Veritas (Abridged)

**Feasibility precedes enforcement** is a structural principle the §5 data illustrates and two 2026 impossibility results motivate (without empirically validating). The Accountability Horizon (Tibebu, 2026) proves no single-locus accountability framework can simultaneously satisfy attributability, foreseeability bound, non-vacuity, and completeness whenever minimum compound autonomy crosses a threshold and the interaction graph contains a mixed feedback cycle; DEMM's *structurally_unfillable* classifications provide a property-level operational analogue. Atomic Decision Boundaries (Fernandez, 2026) proves split evaluation architectures cannot guarantee execution-time admissibility; post-hoc reconstruction classifies whether evidence was sufficient but cannot retroactively establish admissibility. Conventional "enforce at runtime, audit after" framing is operationally vulnerable in cells where evidence is structurally unfillable: there is no audit surface to verify what enforcement produced.

The §5 multi-agent cross-stack cell at 53.6% completeness is concentrated in partial-confidence downgrades and *structurally_unfillable* properties, raising a prima facie auditability risk signal under EU AI Act Article 12 (record-keeping) and Article 14 (traceable oversight) when actor *identity/principal* authority is structurally unfillable in 28.6% overall (European Parliament and Council, 2024). This is a risk signal, not a determination; the EU AI Act specifies no numeric completeness threshold. The 2026 systematic mapping (Nannini et al., 2026) reaches the same conclusion at the instrument level. Adjacent regulatory-instrument literature (Gardhouse et al., 2026; Goncalves, 2026; Beshane, 2026) supplies the cross-jurisdictional comparator. The SR 26-2 banking framework (Federal Reserve System; OCC; FDIC, 2026) and MAS Veritas (Allen et al., 2025) extend the comparison to financial-sector and regional regimes. The method does not establish EU compliance or cover the full regulatory stack.

**Boundary against AI-governance maturity-model literature.** Organisation-level frameworks — AI-RMF-based (Dotan et al., 2024), CMMI-based (Ramdhani & Surendro, 2022), responsible-AI (Papagiannidis et al., 2025), AAGMM (D. Acharya, 2026b) — measure programme maturity at the organisation layer; DEMM operates per-deployment-class per-property. A high organisation-level rating can coexist with $S(Q, P, E)$ failure on a specific audit question, as §1's container fallacy predicts.

## 6.5. Threats to Validity

Thirteen threats to validity (T1–T13) qualify the findings, accompanied by one corpus-composition limitation (L1) recorded outside the threats axis: T1 through T5 are internal to the design, T6 through T9 emerge from the comparative analysis, T10 cedes broader-auditability-framework precedence, T11 records telemetry-execution desynchronisation, T12 records boundary-detection threshold sensitivity, T13 records compensatory property aggregation, and L1 records the preprint-dominant corpus envelope. Each item is detailed below; conclusion §7 maps each T-threat either to an F1–F8 future-work line, to a permanent scope boundary, or to a conclusion-level limitation recap, and carries L1 through to its limitations envelope unchanged.

**T1 — Self-fulfilling synthetic design.** The most serious threat and the reason §5 is framed as feasibility, not validation. The synthetic generator and schema mapper share authorship with the three-break taxonomy paper (Solozobov, 2026e); the mapper encodes architecture-aware downgrade rules that express the hypothesised mechanisms. The severity ordering is hypothesis-consistent by construction. Independent-trace replication with frozen mapper rules (F1) is the precondition for empirical-in-strong-sense status.

**T2 — Named-incident authorship.** §5.4 reconstructions follow the §4.1 *public record establishes / paper infers* convention, not instrumented executions; the 57.1% score measures DEMM on public-report reconstructions, and post-mortems are selected for dramatic failures. Vendor-grade forensic traces (Kent et al., 2006) and AI Incident Database population-scale evaluation (McGregor, 2020) would address the inference step and small-sample sub-threat.

**T3 — Single-instrument evaluation.** All numbers are reconstructor-specific. DEMM evaluates only traces already allowed to happen — no analog of TBA's stress-testing (Paduraru et al., 2026), so §5 reports nominal-execution behaviour. Joint TBA + DEMM adversarial evaluation (Solozobov, 2026f) bridges the gap. Protocol-security analysis (N. Acharya & Gupta, 2026a; Zheng & Zhang, 2026) is a different measurement object at the protocol layer.

**T4 — Architecture-class coverage.** Three architectures × two stack-coverage tiers is a coarse partition; hybrid $planner/executor + supervisor$ patterns (Jenzer, 2026), MedSkillAudit (Hou et al., 2026), and embodied EmbodiedGovBench systems (Qin et al., 2026) sit at sub-coordinates the matrix does not cover.

**T5 — Absence of independent ground-truth verifier.** AER (Vispute & Kadam, 2026) runs a transport-layer reconciler producing a fidelity score against the agent's self-reported record; DEMM has no analogous reconciler and emits feasibility categories from mapper rules. The present paper therefore validates reproducible execution of a published classification protocol, not per-classification truth; §4.3 and §5.7 treat this as a measurement-validity boundary. AER-style integration is F2 of §7.

**T6 — Multi-principal scope limitation.** AgentCity-style cross-organisational agent economies (Ruan & Zhang, 2026a) sit at a coordinate the present 3 × 2 matrix does not cover. Separation-of-power and agent-economy framings (Ruan, 2026b; Kurtz & Krawiecka, 2026) extend the same coordinate. Cross-organisational-trust treatments (Yuan et al., 2026b; Wang et al., 2026) supply parallel anchors. Cross-principal extension is F5 of §7.

**T7 — Absence of assume-compromise evaluation.** DEMM uses nominal synthetic traces and does not test adversarial fragments. Assume-compromise extension inherits Parallax (Fokou, 2026) and Agents-of-Chaos (Shapira et al., 2026) methodologies. The NIST AI $100-2/OWASP\ LLM/MITRE$ ATLAS standards substrate (Vassilev et al., 2025; OWASP Foundation, 2025; MITRE Corporation, 2025) provides the regulatory anchor. F6 of §7 is the follow-up.

**T8 — Information-flow label propagation absent.** Parallax propagates data-sensitivity labels (Fokou, 2026); DEMM's fragment schema lacks sensitivity provenance, so the tensor cannot distinguish reconstructable safe and unsafe actions with the same trace shape. F8 of §7 closes this gap.

**T9 — Atomicity gap.** Atomic Decision Boundaries (Fernandez, 2026) proves split evaluation cannot guarantee execution-time admissibility; DEMM is split post-hoc by design and classifies whether evidence was sufficient for audit, not whether execution was admissible at the time. Permanent scope boundary.

**T10 — Broader-auditability-framework precedence.** Auditable Agents (Nian et al., 2026a) defines the broad framework; the present method is narrower — a property-level sufficiency layer above auditability dimensions and evidence containers (§2.2, §6.6, §6.8). Permanent scope boundary.

**T11 — Telemetry-execution desynchronization.** DEMM classifies *structurally_unfillable* when no record is present (Solozobov, 2026a), conflating (a) governance gaps where evidence was never produced or non-cooperatively stripped with (b) transport-layer failures where execution succeeded but the telemetry write failed. AER illustrates the stronger capture design this paper does not implement (Vispute & Kadam, 2026). §5 measures persisted-evidence sufficiency, not ground-truth-execution sufficiency; differential diagnosis is a logging-layer property outside DEMM's post-hoc scope. Permanent scope boundary.

**T12 — Boundary-detection threshold sensitivity.** §5.2 F1 numbers depend on four configurable heuristic thresholds (§3.3) fixed in the deposited release (Solozobov, 2026a) but not separately calibrated. The multi-agent within-stack F1 = 0.13 may be partly attributable to threshold over-firing on continuous delegation. Sensitivity analysis and recalibration on independent traces (F1 of §7) would separate threshold-induced from structural signal.

**T13 — Compensatory property aggregation.** The completeness metric is a scalar weighted average over the seven implementation properties (Solozobov, 2026a); a high score on one property can offset a zero on another. Real audit or procurement may be non-compensatory for specific prerequisites (e.g., unacceptable if actor *identity/principal* authority is missing). DEMM reports the per-property tensor alongside the scalar; F7 of §7 specifies both aggregate thresholds and non-compensatory per-property minima.

**L1 — Corpus composition.** The 2026 corpus is preprint-dominant; peer-reviewed anchors are Auditable Agents (Nian et al., 2026a) and the synthesis paper (Solozobov, 2026e). L1 sits on the corpus axis rather than the validity-threats axis (T1–T13) and is restated unchanged in §7's limitation envelope.

Taken together, the thirteen threats limit the empirical standing to: hypothesis-consistent outputs on single-principal synthetic scenarios under cooperative-evidence assumptions, deposited heuristic thresholds, and a compensatory scalar reported alongside the per-property tensor; post-hoc reconstruction feasibility, not execution-time admissibility; a narrow property-level sufficiency-assessment layer above auditability dimensions, not a broad auditability framework. Stronger empirical claims require the §7 follow-up work. The envelope follows from the deposited artefact (Solozobov, 2026a), the authorship-overlap dependency (Solozobov, 2026e), and the conceded framework boundary (Nian et al., 2026a).

## 6.6. Governance-Evidence Sub-Taxonomy within the Beyond Task Success Framework

Within Beyond Task Success's four-layer framework (Koch & Wellbrock, 2026), three governance-evidence modes differ by temporal position: runtime defence, contemporaneous capture, and retrospective reconstruction; DEMM is the retrospective instance. AI safety cases (Feakins et al., 2026) supply complementary system-level argumentation. Runtime defence covers TBA (Paduraru et al., 2026), Parallax (Fokou, 2026), and Runtime Governance (Kaptein et al., 2026). Contemporaneous capture is exercised by AER (Vispute & Kadam, 2026). ODTA assigns requirements across modes.

## 6.7. Independent Co-Description of the Three-Break Structure

TBA's failure taxonomy enumerates five operational classes (Paduraru et al., 2026): F1 (coordination collapse and non-termination), F2 (error amplification and unsupported-claim propagation), F3 (role drift and boundary violations), F4 (*tool/memory* injection and interface poisoning), and F5 (misuse outcomes). Of these, F1, F2, and F3 describe the same phenomenon space as the three structural breaks of the synthesis paper (Solozobov, 2026e): F1 corresponds to decision diffusion, F2 to evidence fragmentation, and F3 to responsibility ambiguity. F4 and F5 cover injection and misuse vectors outside the three-break taxonomy and are not load-bearing for the co-description argument. TBA was published in March 2026, the synthesis paper in April 2026; two independent tracks (runtime verification vs post-hoc audit) arrived at overlapping partitions within weeks. The decomposition DEMM's mapper encodes is therefore independently *co-described* — i.e., the same phenomenon space is partitioned along compatible analytic axes by two independent literatures — rather than externally validated; §5 severity ordering remains self-fulfilling under T1 because mapper rules are author-authored, and external validation requires independent labelled traces (F1 of §7).

## 6.8. Landscape Position and Extended Differentiation Triangle

The novelty defence is **three-layered**. Auditable Agents (Nian et al., 2026a) defines auditability dimensions at the framework layer. The substrate layer aggregates surveyed evidence-container literature, and DEMM assesses whether those records suffice for property-level reconstruction at the assessment layer. Three of the substrate regimes cover record-keeping and compliance: PROV-AGENT (Souza et al., 2025), lifecycle audit ledgers (Ojewale et al., 2026), and OSCAL-style evidence (Ugarte et al., 2026). Two cover runtime and reasoning capture: AER (Vispute & Kadam, 2026) and TBA MAT (Paduraru et al., 2026). Two cover delegation and state: IEEC (He & Yu, 2026) and memory-runtime graphs (Lupascu & Lupascu, 2026). Adjacent systems detect, enforce, recover, or present; DEMM occupies the **assess** coordinate.

DEMM is the only framework at the retrospective + non-cooperative + property-level + cross-regime *sufficiency assessment* coordinate within the surveyed window, narrower than Auditable Agents' broad framework (Nian et al., 2026a) and bounded by Beyond Task Success's ODTA Attestability layer (Koch & Wellbrock, 2026) plus the DTR implementation (Solozobov, 2026a). This is a niche-placement finding scoped to the surveyed window, not a dominance claim.

The **container fallacy** diagnostic (§1) is placed against AAGMM (D. Acharya, 2026b), OSCAL-style compliance evidence (Ugarte et al., 2026), and AI Trust OS (Bandara et al., 2026) along five axes — object, output, anchor, audience, and **measurement granularity**; none measures decision-level evidence sufficiency as primary object at the per-deployment-class per-property maturity coordinate. AAGMM measures organisation maturity, OSCAL standardises compliance evidence, AI Trust OS measures telemetry posture, the container-fallacy diagnostic names per-decision sufficiency, and DEMM aggregates that layer into per-deployment-class maturity.

# 7. Conclusion

This paper specified the **Decision Evidence Maturity Model (DEMM)** — a property-level method that operationalises governance-evidence sufficiency for an automated decision as reconstructability over a finite property schema, discrete sufficiency categories, and a five-level maturity rubric, refining Beyond Task Success ODTA Attestability (Koch & Wellbrock, 2026) at the property level over the *Decision Event Schema* (MIT) (Solozobov, 2026b). The contribution is threefold: (i) a per-question, per-property sufficiency relation; (ii) a five-level DEMM maturity rubric anchored to the $CMM/CMMI$ lineage (§3.7); and (iii) the Decision Trace Reconstructor feasibility exercise on 140 synthetic scenarios plus three named-incident reconstructions. The §5 protocol outputs span 53.6-100.0% reconstruction completeness, place agentic cells 24.9 pp below baseline on average, and show a 14.0 pp instrumentation premium for public named incidents. These are fixture-suite outputs, not population estimates. The contribution is narrower than Auditable Agents' broad auditability framework (Nian et al., 2026a): DEMM measures whether records produced under any regime suffice for property-level governance reconstruction, occupying the per-deployment-class per-property maturity coordinate of §6.8.

**Limitations.** The corpus-composition limitation L1 of §6.5 carries through directly: the 2026 corpus is preprint-dominant; peer-reviewed anchors are Auditable Agents (Nian et al., 2026a) and the synthesis paper (Solozobov, 2026e). Four further envelopes recap §6.5's T1–T13 validity threats at the conclusion level. The §5 self-fulfilling synthetic design (recap of T1) is what F1 converts from hypothesis-consistent to empirically grounded. The named-incident reconstructions follow the §4.1 *public record establishes / paper infers* convention rather than vendor-grade forensic traces (recap of T2). Per-class executable conformance is incomplete (recap of T3); evaluation outside the ten adapter-fallback classes waits on F3. The five-level rubric thresholds are proposed design parameters (recap of T13); F7 is the precondition for inheriting values.

**Threat-to-future-work closure map.** The thirteen T-level §6.5 threats close as follows: T1 -> F1; T2 -> §4.1 *public record establishes / paper infers* convention recap; T3 -> F1 plus executable-conformance recap (F3); T4 -> permanent matrix-scope boundary; T5 -> F2; T6 -> F5; T7 -> F6; T8 -> F8; T9 -> permanent post-hoc-scope boundary grounded by Atomic Decision Boundaries (Fernandez, 2026); T10 -> permanent broad-framework-precedence boundary conceded to Auditable Agents (Nian et al., 2026a); T11 -> permanent telemetry-layer boundary; T12 -> F1; T13 -> F7 plus rubric-threshold recap. T4, T9, T10, and T11 are permanent scope boundaries; the other nine map to F1 through F8 future-work, with conclusion-level recaps for T2, T3, and T13 noted in the limitations envelope above. The §6.5 corpus-composition limitation L1 carries through to the §7 limitations envelope unchanged.

Eight future-work lines follow: **F1** independent-trace replication with frozen mapper and T12 sensitivity analysis; **F2** AER reconciler integration (Vispute & Kadam, 2026); **F3** executable conformance for adapter-fallback classes and protocol-level regimes; **F4** *conflicting* scoring and precedence rules; **F5** multi-principal matrix scope, bounded by AgentCity (Ruan & Zhang, 2026a) and separation-of-power framing (Ruan, 2026b); **F6** assume-compromise evaluation via Parallax methodology (Fokou, 2026); **F7** non-compensatory rubric threshold calibration and *Overclaim Rate* measurement (Solozobov, 2026f); and **F8** information-flow label propagation using Parallax's IFL substrate (Fokou, 2026).

**Broader impact and misuse pathways.** Beneficial uses: post-incident audit reconstruction, procurement-time sufficiency screening, citable rubric for academic discourse. Three misuse pathways: **maturity-rating gaming** (reporting high aggregates by selecting property classes the audit won't test) — mitigated by §3.7's per-property minimum-aggregation rule and the DTR's per-property output schema (Solozobov, 2026a); **threshold-shopping** (citing the most permissive threshold from competing models) — mitigated by F7 calibration discipline; **baseline-padding** (deploying multiple containers decoratively and claiming joint presence implies sufficiency) — mitigated by the per-question per-property check of $S(Q, P, E)$ (§1) over the Decision Event Schema (Solozobov, 2026b).